\documentclass[aps,epsfig,twocolumn,showpacs]{revtex4}
\usepackage{graphicx}
\usepackage{dcolumn}
\begin{document}
\title{Momentum distributions of $\alpha$-particles from decaying low-lying 
$^{12}$C-resonances}

\author{R. \'Alvarez-Rodr\'{\i}guez$\:^1$, A.S. Jensen$\:^1$, 
E. Garrido$\:^2$, D.V. Fedorov$\:^1$, H.O.U. Fynbo$\:^1$}
\affiliation{$^1 \:$ 
Department of Physics and Astronomy, University of Aarhus, \\
DK-8000 Aarhus C, Denmark}
\affiliation{$^2 \:$ Instituto de Estructura de la Materia,
Consejo Superior de Investigaciones Cient\'{\i }ficas, \\
Serrano 123, E-28006 Madrid, Spain}

\date{\today}

\begin{abstract}
The complex scaled hyperspherical adiabatic expansion method is 
used to compute momentum and energy distributions of the three
$\alpha$-particles emerging from the decay of low-lying
$^{12}$C-resonances.  The large distance continuum properties of the
wave functions are crucial and must be accurately calculated.  We
discuss separately decays of natural parity states: two $0^+$, one $1^{-}$,
three $2^+$, one $3^-$, two $4^+$, one $6^+$, and one of each of
unnatural parity, $1^{+}$, $2^-$, $3^+$, $4^-$.  The lowest natural
parity state of each $J^{\pi}$ decays predominantly sequentially via
the $^{8}$Be ground state whereas other states including unnatural
parity states predominantly decay directly to the continuum. We
present Dalitz plots and systematic detailed momentum correlations 
of the emerging $\alpha$-particles. 
\end{abstract}

\maketitle

\section{Introduction}

The low-lying resonance states of $^{12}$C have been studied over many
years both theoretically and experimentally, motivated partly by their
astrophysical importance
\cite{mor56,tak70,fri71,ueg77,pic97,des02,nef04,fedt04,kur07,fre07a}. 
Surprisingly, many issues are still not
really understood, e.g. the energies, angular momenta,
structure and decay properties of the resonances. Completely open
questions still remain on the $2^+$ resonances. Morinaga conjectured
in the fifties that a $2^+$ state should exist around 9~MeV as a
member of the rotational band with the $0^+$ resonance at 7.65~MeV as
band-head \cite{mor56}. Several experiments recently provided new
results \cite{joh03,ito04,dig05} but unfortunately no agreement has yet been
reached for the position and width of the first $2^+$ resonance.

Attempts to obtain information about the spectrum from decay
measurements immediately face the problem that only the final state is
observed.  Properties of the initial state must then be reconstructed
from the momentum distributions of the three fragments after the
decay.  Both initial state and the intermediate paths connecting
initial and final states are not observables.  These configurations
can therefore only be described through model interpretations.
This is somewhat different in reaction experiments, where information
can in addition be extracted from properties of outgoing particles
in transfer or scattering reaction.

If we assume that the initial state is a resonance populated one way
or another, and that its decay is independent of the previous history.
This is a simplification decoupling the formation from the decay in
analogy to compound nuclear reactions.  The decay process can then be
viewed as a stationary wave function connecting initial and final
states through a continuous series of intermediate configurations.
This is equivalent to a time dependent process where the initial
state, formed at small distances, evolves through the intermediate
configurations and results in the final state at large distances.
This implies a steady state outgoing flux described precisely by the
stationary resonance wave function.

Thus the resonance wave function can be interpreted as reflecting the
decay mechanisms.  Two principally different modes are traditionally
considered, i.e. sequential decay via an intermediate two-body
configuration, and decay directly into the three-body continuum.  In
both cases the final state is embedded in the three-body continuum and
the modes can only be distinguished if the momentum distributions
carry unique information characterizing one of the modes.  Otherwise
the distinction becomes fluent or a matter of an artificial, although
perhaps more precise, model definition. Previous approaches to
describe this type of observables have been performed mainly for the
$1^+$ states, e.g. Faddeev calculations \cite{tak}, R-matrix
computations, which describe their decay as sequential \cite{bal}, and
Kurchatov fitting, which describes them as direct or democratic
\cite{kor}.

The purpose of this paper is to present $\alpha$-particle momentum
distributions and Dalitz plots \cite{dal53} after decays of all the
computed $^{12}$C-resonances \cite{alv07a} below the proton separation
threshold at an excitation energy of 15.96~MeV, where only
$3\alpha$-decay is possible.  These distributions should help to
establish spins and parities of the yet unknown levels.They provide
then information about structures of initial and intermediate
states. Combined with the measurements a more complete picture of the
$^{12}$C-spectrum and the decay mechanisms should then emerge.  In
section 2 we first sketch the theoretical framework and the choice of
interactions.  The results are presented and discussed in section 3
for both unnatural and natural parity states.  Section 4 contains a
summary and the conclusions.

\section{Theoretical framework}

The resonances decay into three particles, therefore we need a theoretical
tool to describe this three-body continuum structure. We employ the
hyperspherical complex rotated \cite{ho02,fed03} adiabatic expansion
\cite{nie01} in coordinate space to compute bound states and
resonances.  This method is able to deal with several simultaneously
bound and nearly bound two-body states in different subsystems.
Relatively large distances can often be calculated accurately with a
specific choice of basis and partial waves.  The Fourier transform
of the wave function
provides the observable momentum distributions.  The three-body model
consisting of $\alpha$-particles requires interactions which reproduce
energies and scattering properties of the $\alpha$-$\alpha$ system.

\subsection{Practical procedure}

We describe $^{12}$C as a 3$\alpha$-cluster system at all distances.
We use Faddeev equations and solve them in coordinate space using the
adiabatic hyperspherical expansion method \cite{nie01,gar05b,fed03}.
The hyperspherical coordinates consist of the hyperradius $\rho$ and
five generalized angles. The angular Faddeev decomposed wave
functions, $\Phi_{nJM}=\sum_{i=1}^3 \Phi_{nJM}^{(i)}$, are chosen for
each $\rho$ as the eigenfunctions of the angular part of the complex
scaled ($\vec r \rightarrow \vec r \exp(i\theta)$) Faddeev equations
\begin{equation}
(T_\Omega-\lambda_n) \Phi_{nJM}^{(i)}+\frac{2m}{\hbar^2} \rho^2V_i
\Phi_{nJM} = 0 \qquad i=1,2,3\;.
\label{fadeq}
\end{equation}
$T_\Omega$ is the angular part of the kinetic energy operator and
$V_i$ is the potential between particles $j$ and $k$, being
\{$i,j,k$\} a cyclic permutation of \{$1,2,3$\}. The total wave
function, $\Psi^{JM}$, is expanded on the hyper-angular
eigenfunctions, i.e.
\begin{equation}
\Psi^{JM} = \frac{1}{\rho^{5/2}}\sum_n f_n (\rho) \Phi_{nJM}
(\rho,\Omega)\;,
\end{equation}
where the $\rho$-dependent expansion coefficients, $f_n (\rho)$, are
the hyperradial wave functions obtained from the coupled set of
hyperradial equations 
\begin{eqnarray}
\left(-\frac{\partial^2}{\partial \rho^2}+ \frac{15/4}{\rho^2}+
\frac{2m}{\hbar^2} [W_n(\rho)+V_{3b}(\rho)-E]\right) \:f_n(\rho)
&&\nonumber \\ = \sum_ {n^\prime =1}^ \infty \hat
P_{nn^\prime}f_{n^\prime} (\rho) &&\;.
\label{radial}
\end{eqnarray}
$W_n(\rho)$ are the angular eigenvalues of the three-body system
Hamiltonian with fixed $\rho$, $V_{3b}$ is the three-body potential,
$E$ is the three-body energy and $P_{nn^\prime}$ are the non adiabatic
terms. The eigenvalues $W_n(\rho)$ of the angular equations
eq.~(\ref{fadeq}) serve as effective potentials.

In order to obtain the resonances we use the complex scaling method.
According to this method, the energy and width of a resonance state
are associated with the complex eigenvalues of a certain
analytically continued Hamiltonian operator. The appropriate operator
results from the rotation of the position vectors of the ordinary
Hamiltonian into the complex coordinate plane
\begin{equation}
\vec r \to \vec r \: e^{i\:\theta} \qquad \theta > 0, \mathrm{ real} \;. 
\end{equation}
This gives rise to the complex-rotated Hamiltonian
\begin{equation}
H_\theta (\vec r) = H (\vec r \: e^{i\:\theta} )\;.
\end{equation}

The complex energy of a resonance corresponds to a pole in the
momentum-space wave function, while in coordinate-space this form
corresponds to a large-distance asymptotic wave function consisting of
outgoing waves.  In other words, the three-body resonance corresponds
to a complex energy solution $E_0=E_R-i\:E_I$ of the system
(\ref{radial}) with the asymptotic boundary condition of an outgoing
wave in every channel $n$
\begin{equation}
f_n(\rho \to \infty) = C_n\: e^{+i\:\kappa\rho}\;,
\label{boundary}
\end{equation}
where $C_n$ is an asymptotic normalization coefficient and $\kappa =
\sqrt {2mE/\hbar^2}$ is the three-body momentum or the conjugate of
$\rho$. It has been seen that this boundary condition determines that
the scattering matrix has a pole at the complex energy $E_0$, being
$E_R$ the position of the resonance and $\Gamma = 2 E_I$ its width.

The $^{12}$C-resonances are not necessarily of three-body character
even though this by definition must be the case at large distances for
$3\alpha$-decay.  We use the three-body model also at small distances
because, like in Gamow's theory of $\alpha$-decay, the detailed
structure at small distances is not important for the decay properties
which only require the proper description of the emerging three
particles.  We use the three-body short-range potential to adjust the
corresponding small-distance part of the effective potential to
reproduce the correct resonance energies which are all-decisive for
decay properties as evident in the probability for tunneling through a
barrier.

At intermediate distances the three $\alpha$-particles are formed and
the potential has a barrier that determines the partial width of the
resonance.  At large distances the resonance wave functions contain
information about distributions of relative energies between the three
particles after the decay. These properties are connected to the
many-body properties at small distances via preformation factors, as
in $\alpha$-decay. An adjustment of the resonance energy is then
needed. After complex rotation the resonance wave function is
characterized by an exponential fall-off at large distance.  Thus the
crucial information is found in relative sizes of the very small
values, $f_n$, of the resonance at large distances which are very
difficult to compute accurately especially when the Coulomb
interaction is present.

\subsection{Momentum distributions}

The complex scaled coordinate space resonance wave function should be
rotated back to real coordinates and Fourier transformed to provide
the observable momentum distributions.  Unfortunately the
corresponding integral is not convergent and a regularization
procedure has to be applied.  The origin is simply that the resulting
wave function should be a non-normalizable outgoing plane wave at
large distances.  We overcome this problem with the Zeldovic
regularization procedure which is well defined for short-range
interactions \cite{fed03}.  In total this amounts to using the angular
part of the coordinate space wave function at a large hyperradius, but
interpreted as the momentum space wave function.  Inclusion of the
Coulomb interaction is achieved by treating it as an ordinary
potential up to a large value of the hyperradius and then extrapolate
the diagonal parts of the adiabatic wave functions with the
numerically obtained Coulomb and centrifugal potentials.

Two different cases must be treated, i.e. sequential and direct decays
distinguished theoretically by the structure of the adiabatic wave
functions \cite{alv07b}.  Direct decay is characterized by structures
where all particles are far apart and as the hyperradius increases all
distances increase proportionally.  The Zeldovic regularized Fourier
transform of the resonance wave function gives in this case the
momentum distributions \cite{fed04}.

Sequential decay is characterized by a wave function describing a
bound-state like structure of two close-lying particles supplemented
by the third particle far away.  For a complex scaled wave function
such a structure would be that of a two-body resonance, provided the
rotation angle $\theta$ is larger than the angle 
corresponding to the energy and
width of this resonance. These structures approach two-body bound
state configurations as the hyperradius increases.  

However, Fourier transformed and rotated back to the real axis, the
wave function should at large distances approach the description of
the third particle (plane or Coulomb wave) leaving the decaying
resonance which has the given two-body energy and width.  This is two
sequential two-body decays, hence the characterizing notation.  The
resulting momentum distributions cannot be obtained from the rotated
wave function but should instead be calculated from the correct
physical description of two two-body decays.  This results in a
Breit-Wigner distribution for the third particle with a width equal to the
sum of two-body and initial three-body resonance widths peaking around
the energy found by subtracting the two-body from the three-body
energies.

\subsection{Interactions}

The basic ingredients are the two-body interactions,$V_i$, between
particles $j$ and $k$, where \{$i,j,k$\} is a cyclic permutation of
\{$1,2,3$\}.  First $V_i$ must reproduce the low-energy two-body
scattering properties which can be obtained independently for each
partial wave resulting in angular momentum dependent or non-local
interactions.  We rely on the experience gained previously especially
through \cite{alv07a}, and we choose an Ali-Bodmer potential
\cite{ali66} slightly modified in order to reproduce the s-wave
resonance of $^8$Be.  The phase shifts are essentially unchanged and
reproduce $\alpha-\alpha$ scattering data but in order to describe sequential
decays properly the two-body subsystems must also have the correct
energies.  In total we use a potential given as
\begin{eqnarray}
V_{\alpha \alpha} &=&  \left( 125 \hat P_{l=0} + 20 \hat P_{l=2} \right)
e^{-r^2/1.53^2} \nonumber \\
&&- 30.18 e^{-r^2/2.85^2}\;,
\end{eqnarray}
where lengths are in fm and strengths are in MeV. The operators $\hat P_l$
project on angular momentum.

The three-body resonance energy and wave function can now be computed
but the energy usually does not coincide with the measured value. It
may be close, indicating that the three-body structure is nearly
correct.  Then only fine-tuning is needed due to the neglected smaller
three-body effects of polarization or excitations of intrinsic
particle degrees of freedom or off-shell effects.  We emphasize that
only three-body effects are missing since the two-body data already is
reproduced by the phenomenological two-body interactions. We then
correct the energy by including a diagonal three-body short-range
interaction chosen to be Gaussian in hyperradius, i.e. $V_{3b}=
S\exp(-\rho^2/b^2)$.  The structures of the resonances are then
maintained \cite{fed96}.  A larger range corresponding to a third
order power law is not selected as e.g. in \cite{tho00} where it is
used to compensate for the limitation in Hilbert space due to the
hyperharmonic expansion in only one Jacobi coordinate.  Our better
basis confines the three-body interaction to be genuinely of
short-range character.

In the actual parameter choice we prefer to maintain the same values
of $b$ and $S$ for different states with the same angular momentum and
parity $J^{\pi}$ but allow variation with $J^{\pi}$. To see the
systematic behavior we then decided to fix $b=6$~fm corresponding to
the hyperradius obtained when the three alphas are touching in an
equilateral triangle.  The strength $S$ is then adjusted to reproduce
one of the observed resonance energies.  The main dependence is
indirect through the variation of the three-body energy and much
less through the shape of the total potential \cite{alv07a}.
In this way we attempt to separate the effects of the initial 
many-body structure from the symmetries related to the angular 
momentum conservation. The strongest influence is expected from 
Coulomb potentials and centrifugal barriers.

\section{Computed distributions}

We find $^{12}$C-resonances below 15.96~MeV for most angular momenta
$J \leq 6$ and all parities, i.e. two $0^+$, three $2^+$, two $4^+$, and
one of each of $1^\pm$, $2^-$, $3^\pm$, $4^-$ and $6^+$ \cite{alv07a}.
Their structures were described in detail in \cite{alv07a} including
the variation with possible interaction parameters. However, only
small and intermediate distance properties are important for energies,
widths and partial wave decomposition. The final state momentum
distributions after decay arise from the large distance properties
which are much more difficult to determine numerically. 

The procedure is to compute ratios of radial wave functions at large
distances.  This supplies the relative weights on the contributions
from each of the adiabatic wave functions. First we have to remove the
contributions from the wave functions corresponding to population of
two-body resonances.  These fractions must be computed as consecutive
two-body decays and their contributions added to the remaining results
from direct decays which are found by absolute square of the wave
function at a large hyperradius followed by integration over the
unobserved angular variables.  

The asymptotic large-distance behavior should be reached by increasing the
partial waves and the basis size used.  This convergence can be tested
by showing independence of the results with variation of the largest
value of the hyperradius.  Failing the test implies that the basis
size is too small, or contrarily a larger hyperradius can be
compensated by a larger basis producing the same result.  It is then
economical to get stability for a hyperradius and basis as small as
possible.  In most cases we find that the asymptotic behavior is
reached for hyperradii larger than about 60~fm.  There is a small
variation of the distributions from 70 to 100~fm, and we have chosen
80~fm as the value of $\rho$ where the energy distributions are
computed.

The results fall in two groups of natural and unnatural parity states,
e.g. implying that sequential decay through $^8$Be($0^+$) is either
allowed or forbidden by conservation of angular momentum and parity.
Decay through $^8$Be($2^+$) is possible in both cases but this state
is rather broad and the result would be hard to distinguish from
direct decay. We see no indication of population of this channel in
the numerical results.  To optimize the accuracy we then maintain as
small a scaling angle as possible consistent with distinct separation of the
three-body resonance from the background continuum.

\subsection{Unnatural parity states}

\begin{figure*}
\vspace*{0.1cm}
\begin{center}
\includegraphics[width=18cm]{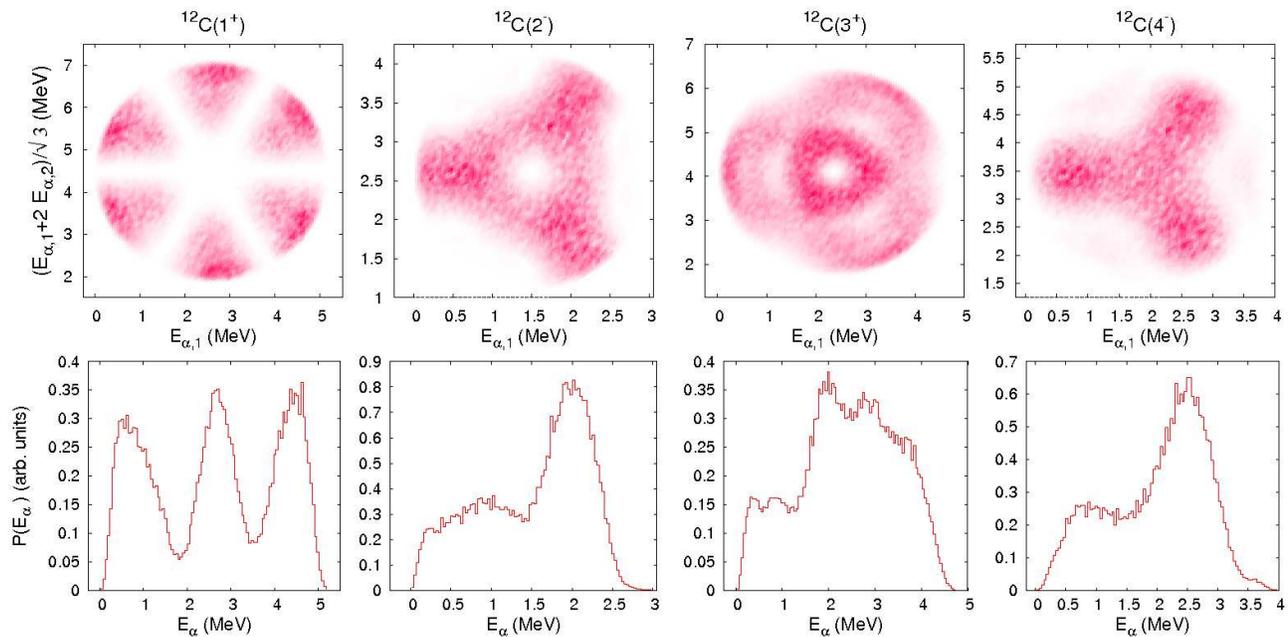}
\end{center}
\caption{Dalitz plot (upper part) and the $\alpha$-particle energy
  distribution (lower part) for the $(1^+, 2^-, 3^+, 4^-)$-resonances
  at an excitation energy of (14.98, 11.80, 14.40, 13.26)~MeV or
  (7.70, 4.53, 7.13, 5.98)~MeV above the 3$\alpha$ threshold, which is
  7.275~MeV above the ground state. We have performed a Monte Carlo
  integration over the phase space, which due to the statistical
  nature produces the unphysical fluctuations.}
\label{fig1}
\end{figure*}

\begin{figure*}
\vspace*{0.1cm}
\begin{center}
\includegraphics[width=14cm,angle=-90]{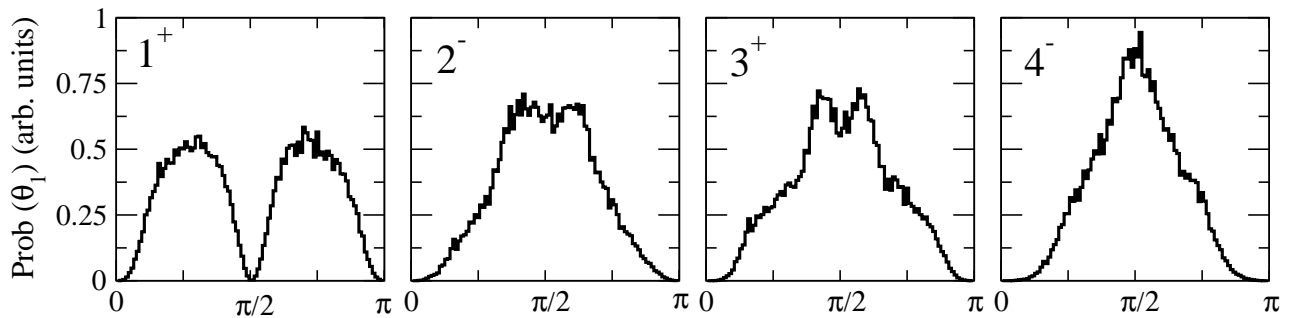}
\end{center}
\vspace*{-9cm}
  \caption{The angular distributions of the directions between two
  particles and their center of mass and the third particle for the
  $(1^+, 2^-, 3^+, 4^-)$-resonances in Fig.\ref{fig1}.  We have
  performed a Monte Carlo integration over the phase space, which due
  to the statistical nature produces the unphysical fluctuations. }
\label{fig2}
\end{figure*}

\begin{figure*}
\vspace*{0.1cm}
\includegraphics[width=14cm,angle=-90]{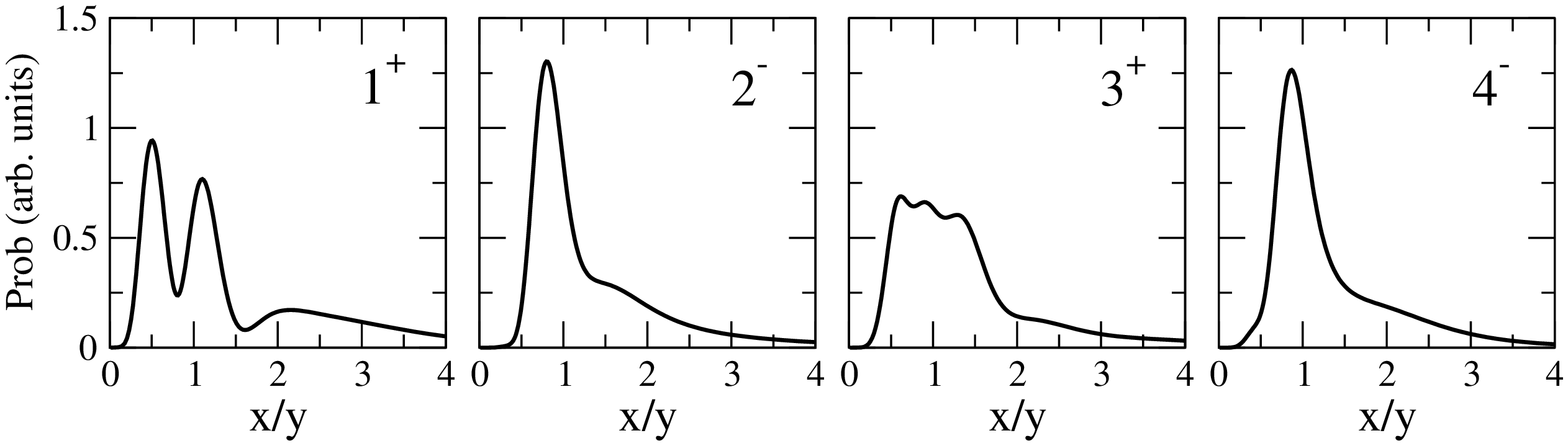}
\vspace*{-9cm}
  \caption{The distributions of the ratio of the distances between two
  particles and their center of mass and the third particle for the
  $(1^+, 2^-, 3^+, 4^-)$-resonances in Fig.\ref{fig1}. }
\label{fig3}
\end{figure*}

These states are $1^+$, $2^-$, $4^-$ and $3^+$ and our basis describes
them as decaying directly although analyses of measured distributions
employ interpretations as sequential through $^8$Be($2^+$)
\cite{dig05,dig06,fyn03}.

The lowest $1^+$ state was briefly discussed previously in
\cite{alv07b,alv07d}.  Experimentally two $1^+$ states, isospin 0 and 1,
are known but we find only one reflecting that we are confined to
isospin 0 by using $\alpha$-particles as building blocks.  Both states
are very far from resembling $\alpha$-cluster states.  Still the
decays of both states must proceed through the same $\alpha$-cluster
configurations, although the weights on the adiabatic potentials might
differ from state to state.  Underlying many-body effects are beyond
the present model but we can pinpoint the neglected effects, i.e. the
preformation factors established at small distances where the
many-body problem is constraint into a three-body problem, and better
three-body potential to account for the transition between the $N$- and
three-body degrees of freedom at short and large distances,
respectively.  For these reasons the contributions from the individual
adiabatic potentials could differ for decays of these two $1^+$ states
of different isospin.

We first focus on the isospin 1 state at an excitation energy of
14.98~MeV ($7.70$~MeV above threshold).  We adjust the three-body
potential and compute the energy distributions shown in
Fig.\ref{fig1}.  The upper part exhibits the Dalitz plot and the lower
part projects the distribution on the axis with one $\alpha$-particle
energy.  The latter is computed by using Monte Carlo integration over
all phase space directly from the wave function.

The measured distributions \cite{bal74} are very uncertain first of
all because the lower-lying isospin zero $1^+$ state at 12.70~MeV
(5.42~MeV above threshold) also is populated via feeding from a gamma
transition between the two $1^+$ states.  This contribution is not
easily removed from the existing data to allow a clean comparison.  A
better analysis or a new experiment measuring the $\alpha$-decay of
the T=1 $1^+$ state in complete kinematics is required.  Our computed
result is almost identical to the distributions, measured and
calculated, for the 12.70~MeV state \cite{fyn03,alv07b} if the
difference in available energy is corrected for. The distributions in
Fig.\ref{fig1} are then direct prediction based on the assumptions
that the isospin zero components in both states are equally populated
and decay through the same mechanism.

A test of this prediction would provide interesting information about
the dynamics of isospin mixing.  Two extremes can be imagined,
i.e. the same isospin $0$ components can be present from small to
large distance resulting in the same distribution, or different
complicated many-body structures at small distances clusterize into
$\alpha$-particles around the nuclear surface and proceed to detection
at large distances.  We know that the partial decay widths for both
states are much smaller than predicted from the cluster model
\cite{alv07a} but this information does not prohibit the momentum
distributions from being almost identical.

In the computation we find only one of each state of even $J$ and
negative parity, i.e. one $2^-$ one $4^-$ state.  The experimental
spectrum has two states of $2^-$ where the highest at 13.26~MeV
(5.98~MeV above threshold) only is tentatively assigned to have $2^-$
\cite{azj} while no $4^-$ state is found experimentally.  It is then
tempting to believe that this state at 13.26~MeV really is a $4^-$
state as indicated by our computations \cite{alv07c}.  This new
spin-parity assignment has also been suggested recently in
\cite{fre07}.

One way to decide which spin and parity is correct is to measure the
momentum distributions of the fragments emerging after decay.  Usually
this carries distinct signatures of the angular momentum of the
decaying state. In \cite{jac72,ant75} it is shown that, even 
within non-sophisticated theoretical models, the basic signatures
of the angular momentum are present in the experimental data.
We first turn to the energy distributions in
Fig.\ref{fig1}.  Both $2^-$ and $4^-$ show very similar distributions
but the peaks appear at slightly higher values for the $4^-$ state.
However, the two-dimensional Dalitz plots differ more from each other.
Both have the triangular symmetry but the $2^-$ resonance have
virtually nothing in between these peaks in contrast to the somewhat
more smeared out distributions of the $4^-$ resonance.

In the computations we find also a $3^+$ resonance for which there is
no experimental evidence, but it has been suggested in \cite{fre07} to
assign these quantum numbers to the state at about an excitation
energy of 13.35~MeV. Theoretically a $3^+$ state has also been found
in \cite{ueg77}.  With a reasonable three-body strength, -20~MeV,
placing the state at 14.40~MeV (7.13~MeV above threshold), we find the
energy distributions in fig.~\ref{fig1}.  The distribution is very
broad but peaked at intermediate energies.  This is seen to arise from
a Dalitz plot distribution with a small hole in the middle surrounded
by a close-lying dense circle and a much larger diffuse distribution.

The angular momentum may leave an even more distinct signature in the
angular distributions, shown in Fig.\ref{fig2}, of the directions
between two particles and their center of mass and the third particle.
The information is then directly about the corresponding angular
momentum denoted as $\ell_y$, i.e.  the angular momentum of the third
particle relative to the center of mass of the other two with the
relative angular momentum $\ell_x$.  We see that the angular
distribution patterns are quite different for different states. The
$1^+$ distribution shows two broad peaks separated by a minimum with
vanishing probability at an angle of $\pi/2$.

This reflects that the partial wave components in the angular wave
function are a linear combination of only $(\ell_x, \ell_y) = (2,2),
(4,4)$ each coupled to the resulting value of $1$ \cite{alv07b}.
Choosing the specific directional angles of $\phi_x = \phi_y =
\theta_x=0$ and $\theta_y = \pi/2$ we find that only projection
quantum numbers of $m_x =0$ and $m_y =0, \pm 2, \pm 4$ give
non-vanishing contributions. This is only consistent with a projection
of the total angular momentum $M= m_x + m_y$ since $M=m_y=\pm 1$ gives
zero. However when all projections are zero the Clebsch-Gordan
coupling coefficient is also zero.  The observables in
Fig.\ref{fig2} reveal information about the intrinsic angular
momenta used to construct the wave function.

In contrast both $2^-$, $3^+$ and $4^-$ have peaks in the
distributions at $\pi/2$.  The different shapes can be traced back to
the different partial wave decomposition computed and discussed in
\cite{alv07b}, i.e. $2^-$ has about 40\% to 60\% of $\ell_y =1,3$,
while $4^-$ is dominated by $\ell_y =3$, and $3^+$ has about twice as much
$\ell_y =2$ as $\ell_y =4$.  These features are clearly
distinguishable demonstrating that these observables can be used to
determine the large-distance structure of these resonances.  The
initial state can still only be determined through the theoretical
information about the dynamical evolution of the resonances.

The one-dimensional distributions in Fig.\ref{fig1} can be used to
extract the distributions of how far the three particles are from each
other \cite{alv08a}.  This is visualized by a triangle with a particle
in each corner moving apart from their common center of mass. In
particular the distributions of the ratio, $x/y$, of the distances
between two particles, $x$, and their center of mass and the third
particle, $y$, are shown in Fig.\ref{fig3}.  Since all Jacobi systems
are identical we do not have to distinguish between Jacobi sets.
Unfortunately the symmetric wave function then do not allow
distinction between these identical particles.

With several peaks as for the $1^+$ resonance the interpretation is
obvious, namely that each peak contains one $\alpha$-particle.  The
triangular geometric structure for the decay of this isospin one 
$1^+$ state then corresponds
to side ratios of 2.2:1.8:1 of the triangle.  For the other unnatural
parity states only one broad peak is seen close to the value 1. For an
equilateral triangle the $x/y$-ratio is $2/\sqrt{3} \approx 1.15$
which then is the only value where a narrow peak is possible.
Otherwise a broader peak must cover overlapping distributions
deviating somewhat from the equilateral triangle and corresponding to
similar but less symmetric configurations.

\subsection{Natural parity states}

\begin{figure*}
\vspace*{0.1cm}
\begin{center}
\includegraphics[width=18cm]{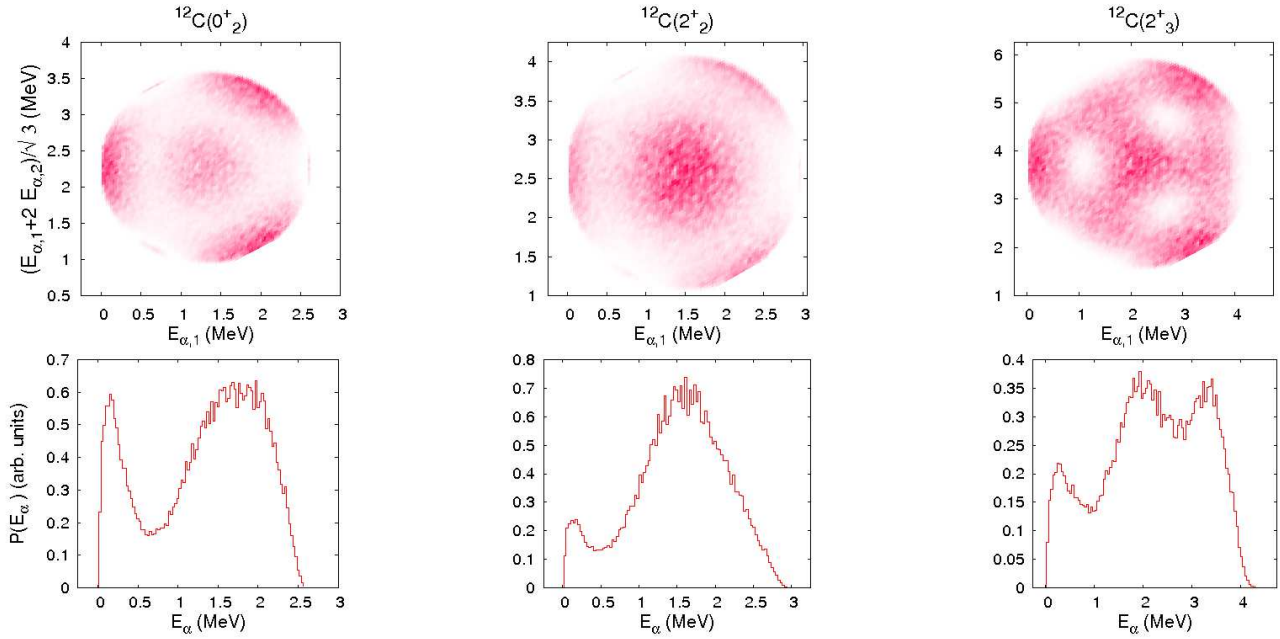}
\end{center}
  \caption{Dalitz plot (upper part) and the $\alpha$-particle energy
  distribution (lower part) for the $(0^+_2, 2_2^+, 2_3^+)$-resonances
  at an excitation energy of (11.22, 11.76, 13.76)~MeV or (3.95, 4.48,
  6.49)~MeV above the 3$\alpha$ threshold, which is 7.275~MeV above
  the ground state.  We have performed a Monte Carlo integration over
  the phase space.  The sequential decay, respectively of 59\%, 15\%,
  4\%, through $^8$Be($0^+$) is removed. We label as in table
  \ref{tab1}.}
\label{fig4}
\end{figure*}

\begin{figure*}
\vspace*{0.1cm}
\begin{center}
\includegraphics[width=18cm]{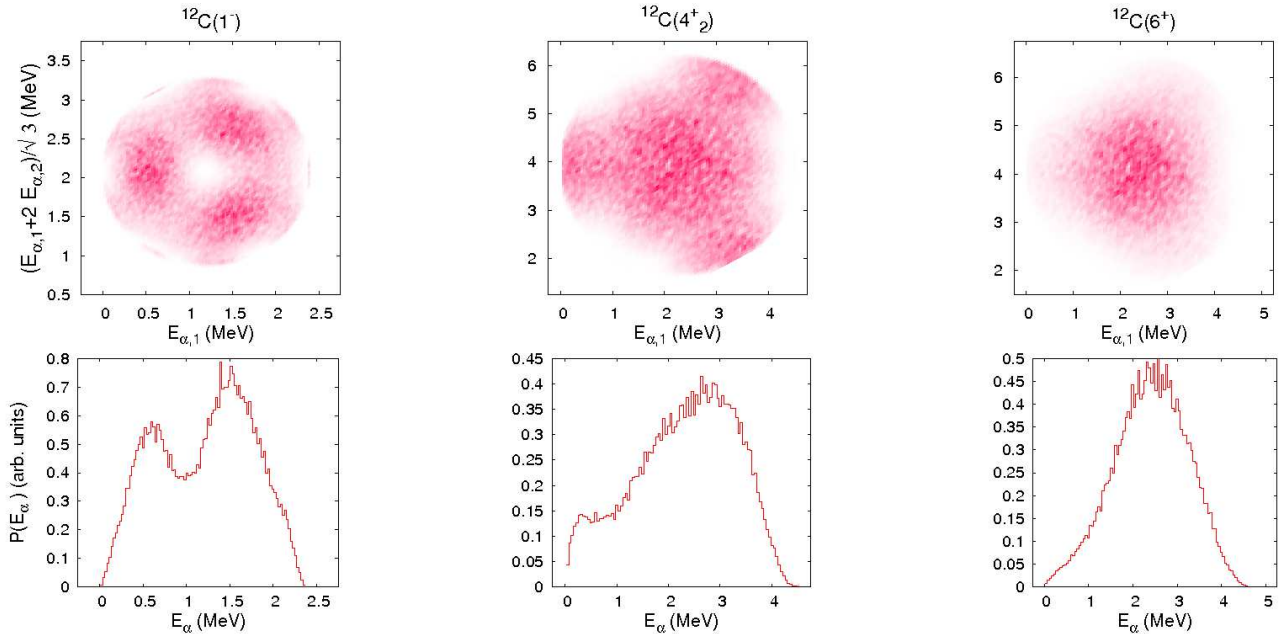}
\end{center}
  \caption{Dalitz plot (upper part) and the $\alpha$-particle energy
  distribution (lower part) for the $(1^-, 4_2^+, 6^+)$-resonances at
  an excitation energy of (10.88, 14.10, 14.40)~MeV or (3.61, 6.83,
  7.13)~MeV above the 3$\alpha$ threshold, which is 7.275~MeV above
  the ground state.  We have performed a Monte Carlo integration over
  the phase space.  The sequential decay, respectively of 70\%, 20\%,
  5\%, through $^8$Be($0^+$) is removed. We label as in table
  \ref{tab1}.
}
\label{fig5}
\end{figure*}

These states are $0^+$, $1^-$, $2^+$, $3^-$, $4^+$ and $6^+$.  They
can decay via the energetically favorable $^8$Be($0^+$) which
asymptotically must be described by one of the adiabatic potentials
with the $^8$Be+$\alpha$ structure.  The signature is simply that this
potential approaches the complex energy of the $^8$Be($0^+$)
resonance. The radial resonance wave functions at large distances
determine the population fractions for each of the adiabatic potentials.  In
particular we can find the fraction of decay proceeding sequentially
through this $0^+$-state, and furthermore we can compute the related
distributions as two consecutive two-body decays.

The result is one peak close to an energy of $E_{max}=2E_{\alpha}/3$
with a width roughly equal to the width of the decaying state, and a
broader square-like peak at an energy of about $E_{max}/4$ determined
by kinematics.  Here we assumed vanishing energy and width of
$^8$Be($0^+$), otherwise the peak positions and widths should be
modified.  The Dalitz plots should also reflect these features by
showing one high-energy, almost vertical, single-$\alpha$
distribution, and two separated (for each of the other
$\alpha$-particles) more horizontal distributions corresponding to a
broader peak after projection on the single $\alpha$-energy $x$-axis.

The angular distribution from sequential decay through $^8$Be($0^+$)
must reflect the behavior of the angular momentum $\ell_y$ precisely
as for ordinary decays of a quantum state of given angular momentum.
The direct decay is expected to give a relatively broad distribution
shifted from the central value at half the maximum energy by an
appropriate average over the combinations of angular momentum phase
space factors.  This can also be interpreted geometrically as an
expanding triangular configuration with given side ratios.

\begin{table}
\vspace*{-0.1cm}
\caption{\label{tab1} Energy above the triple-$\alpha$ threshold,
excitation energy and estimated amount of sequential via $^8$Be($0^+$)
and direct decays for the natural parity states of $^{12}$C. If
necessary we label the resonances with increasing energy above
threshold. }
\begin{footnotesize}
\begin{center}
\begin{tabular}{|ccccc|}
\hline
$J^\pi$& E$_{\alpha \alpha \alpha}$ (MeV) & E$_{exc}$ (MeV) & 
sequential  & direct  \\
\hline
$0^+_1$ & 0.38 & 7.66 &95\% & 5\%\\
$2^+_1$ & 1.38 & 8.66 &97\% & 3\%\\
$3^-$   & 2.33 & 9.60 &96\% & 4\%\\
$4^+_1$ & 3.25 & 10.52&92\% & 8\%\\
$1^-$   & 3.61 & 10.88&70\% & 30\%\\
$0^+_2$ & 3.95 & 11.22&59\% & 41\%\\
$2^+_2$ & 4.48 & 11.76&15\% & 85\%\\
$2^+_3$ & 6.49 & 13.76& 4\% & 96\%\\
$4^+_2$ & 6.83 & 14.10&20\% & 80\%\\
$6^+$   & 7.13 & 14.40& 5\% & 95\%\\
\hline
\end{tabular}
\end{center}
\end{footnotesize}
\end{table}

First we extract the percentage of sequential decay via $^8$Be($0^+$)
and direct decay for the natural parity states, see table \ref{tab1}.
The lowest-lying natural parity states of each $J^{\pi}$ ($0^+$,
$2^+$, $3^-$ and $4^+$ states with excitation energies 7.66~MeV,
8.66~MeV, 9.60~MeV and 10.52~MeV) seem to be completely dominated by
decays via $^8$Be($0^+$).  In contrast, the highest-lying $2^+$ state
at 13.76~MeV excitation energy and the $6^+$ state at 14.40~MeV
excitation energy only have small fractions decaying through the
$^8$Be ground state.  In the remaining cases ($1^-$, $0^+$, $2^+$ and
$4^+$ states with excitation energies 10.88~MeV, 11.22~MeV, 11.76~MeV
and 14.10~MeV) both mechanisms are comparable.

The lowest of the two $0^+$ resonances is the so-called Hoyle state,
which plays an important role in nuclear astrophysics. According to
our computation, it decays almost entirely sequentially.  Very little
is left for the direct decay which therefore is not shown.  The
experimental distribution is also consistent with complete domination
of sequential decay as in our computation \cite{alv07b}.  We also omit
the other three natural parity resonances dominated by sequential
decays.  We concentrate instead on the 6 resonances where a
substantial amount is direct decay.  These distributions are shown in
Figs. \ref{fig4} and \ref{fig5} after removal of the contributions
from the sequential decay through the $^8$Be ground state.  The
experimental analyses can extract the trivial contribution from the
decay through the $^8$Be ground state.  It is therefore
straightforward to make a comparison with the experiment.  Both Dalitz
plots and projected single-$\alpha$ energy distributions are shown.

The higher-lying $0^+$ resonance has a large width of about 3.5~MeV.
On top of this difficulty the population through beta-decay of the
corresponding energy region leads to violation of the independent
approximation of formation and decay of the resonance. The main effect
is a shift in energy of the resonance position. In any case this state
has a significant probability of decaying directly into the three body
continuum.  This part, shown in fig.~\ref{fig4}, exhibits a triangular
structure in the Dalitz plot, but now we find one low-energy
$\alpha$-particle and two of moderate energies.  This is in almost
complete contrast to the sequential decay where one energy is high and
two are small.

Next we focus on our results in connection with the existence and
position of low-lying $2^+$ resonances which still is an open question
for the $^{12}$C nucleus.  The old suggestion is that the Hoyle state
should be the band-head followed by a $2^+$ state at around 10~MeV
\cite{mor56}.  There are experimental indications for the existence of
such a state \cite{fyn03} but no consensus has so far been reached.
On the other hand other theoretical models, also cluster models, find
three $2^+$ resonances in this energy region \cite{kan06}; in
\cite{ueg77} two $2^+$ excited states are found in this region, while
in \cite{des02} one $2^+$ state appears below 12.3~MeV excitation
energy.  We find rather different structures for these three states,
still all of $\alpha$-cluster structure. Each of them is dominated by
its own adiabatic wave function corresponding to three different
low-lying adiabatic potentials with differing partial wave
decomposition \cite{alv07a}.  Most likely these states are hidden
behind broad states of roughly the same energy.  They are therefore
extremely difficult to distinguish from the background in any of the
experiments.

Their decay properties also vary substantially, e.g. the percentage of
sequential decay through $^8$Be($0^+$), see table \ref{tab1}.  The
lowest state almost exclusively decays sequentially while the other
two mostly decay directly.  In fig.~\ref{fig4} we see that the direct
parts give very broad distributions.  For the second $2^+$ resonance
all three $\alpha$-particles emerge with large probability with
similar kinetic energies.  For the third $2^+$ resonance the
distribution is more diffuse and the energies are more unevenly
divided resulting in a structured but relatively broad distribution.

We continue with the $1^-$ state, both Dalitz plot and
one-dimensional projection are shown in Fig.~\ref{fig5} for the 30\%
decaying directly into the three-body continuum.  A similar triangular
structure as for the second $0^+$ resonance is seen although
substantially more smeared out resulting in two overlapping broad
peaks after projection on the $x$-axis.

Both the $3^-$ resonance and the lowest of the two $4^+$ resonances
are almost completely dominated by sequential decay.  The second of
the $4^+$ resonances gives a rather diffuse distribution of kinetic
energy of the $\alpha$-particles, see fig.~\ref{fig5}.  It resembles
somewhat the distribution from the third $2^+$ resonance except 
that the small probability holes in the Dalitz plot now also are
smeared out.  This distribution is again almost the opposite of the
sequential decay distribution with one high and two low energy
particles.  The $6^+$ resonance has a symmetric distribution extending
about 1~MeV around a central region where all energies are roughly
equal.

\begin{figure}
\vspace*{0.1cm}
\includegraphics[width=8.5cm]{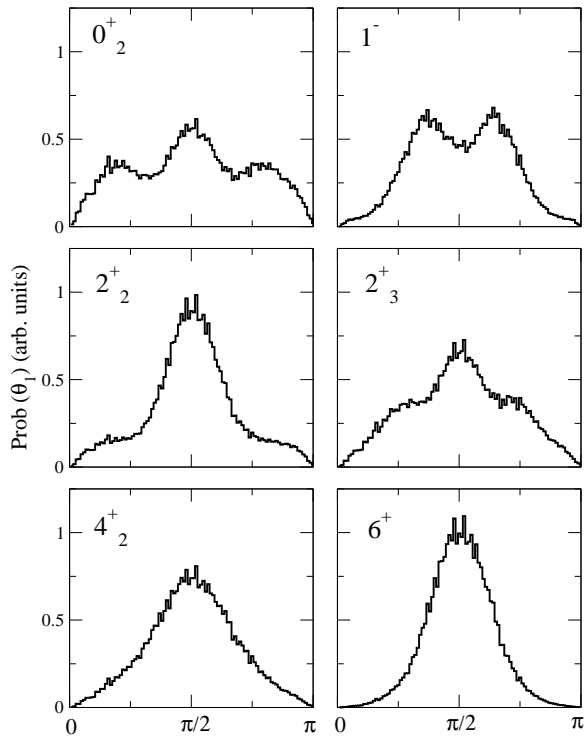}
  \caption{The angular distributions of the directions between two
  particles and their center of mass and the third particle for the
  $(0_2^+, 1^-, 2_2^+, 2_3^+, 4_2^+, 6^+)$-resonances in
  Figs.\ref{fig4} and \ref{fig5}.
  We have performed a Monte Carlo integration over the phase space.
  The sequential part is removed as in figs. \ref{fig4} and \ref{fig5}.
  We label as in table \ref{tab1}.}
\label{fig6}
\end{figure}

\begin{figure}
\vspace*{0.1cm}
\includegraphics[width=8.5cm]{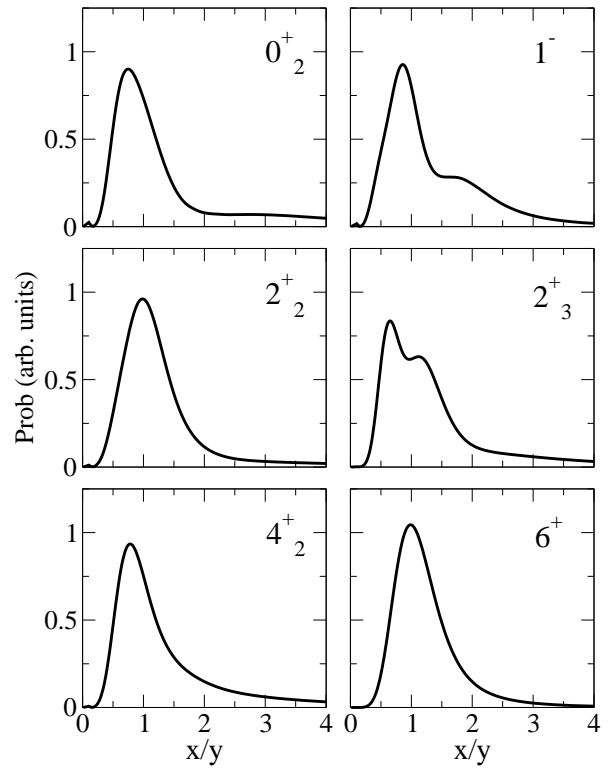}
  \caption{The distributions of the ratio of the distances between two
  particles and their center of mass and the third particle for the
  $(0_2^+, 1^-, 2_2^+, 2_3^+, 4_2^+, 6^+)$-resonances in
  Fig.\ref{fig4}.
  The sequential part is removed as in figs. \ref{fig4} and \ref{fig5}.
  We label as in table \ref{tab1}.}
\label{fig7}
\end{figure}

We now turn to the other type of information found in the angular
distributions which exhibit the correlated directions of emergence.
Obviously the sequential decay through the $^8$Be($0^+$) state must be
with the third $\alpha$-particle in the opposite directions of $^8$Be.
The only information here is then about the partial wave component
($\ell_y$) of that third particle relative to $^8$Be.  Angular
momentum conservation then requires the total angular momentum $J =
\ell_y$.  Thus the most interesting new information is contained in
the directly decaying parts shown in Fig.\ref{fig6}.  These
distributions also vary from state to state reflecting the structure
in terms of partial waves as discussed in \cite{alv07a}.

The distribution corresponding to $0_2^+$-state is essentially from
the isotropic distribution of $\ell_y=0$ modified by a smaller
contribution from $\ell_y=2$ with maxima at $\pi/4$ and $3\pi/4$
separated by zero probability at $\pi/2$.  The distribution
corresponding to $1^-$ shows two peaks separated by a small minimum at
$\pi/2$.  The largest partial waves are here $\ell_y=1,3$.  The
angular distributions of both the second and third $2^+$ resonance
seem to contain a narrow peak on top of a broader one.  These
structures are due to large contributions from $\ell_y=0$ supplemented
by contributions from $\ell_y=2$ and $\ell_y=4$, respectively.
Finally, the distributions form $4^+_2$ and $6^+$ both exhibit one
smooth, and for $6^+$ also relatively narrow, peak around $\pi$/2.  
The partial wave structures of
these states are mainly $\ell_y=2$, and $\ell_y=2,4$, respectively

We again attempt to extract the geometric structure of the dominating
triangular decay configurations.  The results for the ratio between
one pair of particles and their center of mass and the third particle
are shown in Fig.\ref{fig7}.  They are all rather similar with a
relatively broad peak around 1, but for $1^-$ and $2_3^+$ with more
structure at a larger ratio suggesting another peak.  As in
Fig.\ref{fig3} the peaks must cover overlapping distributions to
correspond to an almost equilateral triangle.  
In the case of $0^+$ a very broad peak appears around 3, and the other
two peaks are around 0.8. This gives rise to an obtuse triangle
with side ratios 1.7:1:1.

\section{Summary and conclusions}

We have computed the $\alpha$-particle momentum distributions of 14
three-body decaying low-lying $^{12}$C many-body resonances with 10
different angular momenta and parities.  The results are exhibited as
single $\alpha$ energy distributions and energy correlations of Dalitz
plots.  We assume that the decays of the resonances are independent of
their formation as for compound nuclear reactions.  We use a
three-$\alpha$ cluster model to describe all states even at small
distances where the cluster model sometimes fails badly and the
many-body structure is indispensable for a structure computation.  The
idea is, the same as for the classical $\alpha$-emission, that three
$\alpha$-particles must be formed at small or intermediate distances
as they emerge at large distances after the decay.  Thus the small
distance properties should only supply boundary conditions and impose
energy and angular momentum conservation.  This we mock up in the
$3\alpha$ cluster model by a three-body interaction adjusted to
reproduce the resonance energy.  Again a simple analogy is found in
the preformation factors in $\alpha$-emission. 

An extreme example is the isospin 1 state which cannot be formed
by $\alpha$-clusters.  Its $\alpha$-decay width is consequently
very small but still the resulting distributions are with the present
assumptions predicted to be essentially the same as the $1^+$ isospin 0
state.

For three-body decays the interest, and the complication, is how the
energy is shared between the three particles.  This is determined by
the ``dynamic evolution'' of the resonances, i.e. by the change in
structure from small to large distances.  To a large extent the
decisive properties are symmetries from angular momentum and parity
conservation.  The resulting momentum distributions carry information
about both initial resonance state and the intermediate configurations
(decay mechanisms).  The only energetically allowed two-body structure
is the ground state of $^8$Be.  Sequential decay through this state is
dominating for natural parity states for the lowest resonance of a
given angular momentum.  The momentum distributions for the fractions
decaying directly are predicted for all resonances below the proton
separation threshold.

Whenever possible we give a geometric description of the parts
decaying directly to the three-body continuum.  This is expressed as
side ratios of the $\alpha$-particles emerging in a triangle.

The Dalitz plots and $\alpha$-energy distributions differ from state
to state.  A complementary observable is the correlation between the
direction of one particle and the center of mass of the other two.
These distributions could be used to assign spin and parity to these
decaying states as soon as sufficiently accurate experimental data
become available.  The directly measured angular distribution must
contain information about the angular momentum of one particle with
respect to the center of mass of the other two particles at large
distances. Since several partial waves may contribute this information 
is not unique, and may have to be supplemented with other information.  
Furthermore, the uncertainty remains of how the measured large-distance 
properties reflect the small and intermediate-distance structures of the
resonance wave function.  Only a theoretical model can provide this
connection.

In conclusion, we provide systematic and detailed decay information 
(fraction of
sequential decay, Dalitz plots, single-$\alpha$ energy distributions,
momentum direction correlations), which can be compared to 
upcoming experimental data, for each of the 14 lowest $^{12}$C
resonances decaying by $3\alpha$-emission.

\begin{center}
{\Large \bf Acknowledgments} 
\end{center}

R.A.R. acknowledges support by a post-doctoral fellowship from
Ministerio de Educaci\'on y Ciencia (Spain).

\end{document}